\documentclass{article}
\usepackage{amsmath}
\usepackage{xspace}

\newcommand{\Id}[1]{\ensuremath{\text{{\sf #1}}}}

\newcommand{\set}[1]{\left\{ #1\right\}}
\newcommand{\gilt}{:}

\newcommand{\setGilt}[2]{\left\{ #1\gilt #2\right\}}

\newcommand{\etal}{et al.\xspace}

\newcommand{\realrange}[2]{\left[#1, #2\right]}

\newcommand{\unitrange}[2]{\realrange{0}{1}}

\newcommand{\Oh}[1]{\mathrm{O}\!\left( #1\right)}

\newcommand{\llabel}[1]{\label{\labelprefix:#1}}
\newcommand{\labelprefix}{} 

\newcommand{\discussionsize}{\small}

\newcommand{\notiz}[1]{}
\newcommand{\frage}[1]{}

\newcommand{\punkt}{\enspace .}

\newenvironment{code}{\noindent
\begin{tabbing}%
\hspace{2em}\=\hspace{2em}\=\hspace{2em}\=\hspace{2em}\=\hspace{2em}\=%
\hspace{2em}\=\hspace{2em}\=\hspace{2em}\=\hspace{2em}\=\hspace{2em}\=%
\kill}{\end{tabbing}}

\newcommand{\labelcommand}{}
\newcommand{\captiontext}{}
\newsavebox{\codeparam}
\newcounter{lineNumber}
\newenvironment{disscodepos}[3]{%
\renewcommand{\labelcommand}{#2}%
\renewcommand{\captiontext}{#3}%
\sbox{\codeparam}{\parbox{\textwidth}{#3}}%
\begin{figure}[#1]\begin{center}\begin{code}\setcounter{lineNumber}{1}}{%
\end{code}\end{center}\caption{\llabel{\labelcommand}\captiontext}\end{figure}}

{\end{disscodepos}}

\newcommand{\Procedure}{{\bf Procedure\ }}

\newcommand{\Repeat}   {{\bf repeat\ }}
\newcommand{\Until}    {{\bf until\ }}

\newcommand{\Do}       {{\bf do\ }}

\newcommand{\Foreach}      {{\bf foreach\ }}

\newcommand{\Is}{\mbox{\rm := }}

\newcommand{\If}       {{\bf if\ }}

\newcommand{\Then}     {{\bf then\ }}

\newcommand{\RRem}[1]   {\`{\bf //\hspace{0.5mm}~}{\rm#1}}

\newdimen\endofsize\endofsize=0.5em
\def\endofbeweis{~\quad\hglue\hsize minus\hsize
                 \hbox{\vrule height \endofsize width
\endofsize}\par}

\newcommand{\ignore}[1]{}

\renewcommand{\frage}[1]{[{\sf#1}]}

\usepackage{multirow}
\newtheorem{theorem}{Theorem}

\DeclareMathOperator{\towns}{towns}
\DeclareMathOperator{\ed}{editDistance}

\DeclareMathOperator{\simil}{sim}
\DeclareMathOperator{\tokens}{tokens}

\DeclareMathOperator{\rating}{rating}
\DeclareMathOperator{\idf}{IDF}

\newcommand{\dk}[1]{\todo[inline]{DK: #1}}

\newcommand{\myText}[1]{``{\tt #1}''}
\newcommand{\rlower}{\underline\rho}
\newcommand{\rupper}{\overline\rho}
\newcommand{\CS}{C_{\mathcal{S}}}
\newcommand{\NeG}{D^*}
\newcommand{\CT}{C_{\mathcal{T}}}
\newcommand{\CST}{C_{\mathcal{T\times S}}}

\usepackage{amssymb}
\usepackage{todonotes}
\usepackage{textcomp}
\usepackage[ngerman,english]{babel}
\newtheorem{proof}{Proof}

\usepackage{fullpage}
\renewcommand{\frage}[1]{\todo[inline]{PS: #1} }
 \renewcommand{\frage}[1]{}
 \renewcommand{\todo}[2][inline]{}
\usepackage[pdfpagelabels=false]{hyperref}
\begin{document}

\title{Efficient Error-Correcting Geocoding
\thanks{Partially supported by DFG Grant 933/5-1.}}
\author{Christian Jung, Daniel Karch, Sebastian Knopp, Dennis Luxen, and Peter Sanders\\\normalsize
Karlsruhe Institute of Technology, 76128 Karlsruhe, Germany\\\normalsize {\tt
\{christian.jung,sebastian.knopp\}, karch@math.tu-berlin.de  \{luxen,sanders\}@kit.edu }}

\maketitle
\begin{abstract}
We study the problem of resolving a perhaps misspelled address of a location into geographic coordinates of latitude and longitude.
Our data structure solves this problem within a few milliseconds even for misspelled and fragmentary queries. Compared to major geographic search engines such as Google or Bing we achieve results of significantly better quality.
\end{abstract}
 



\section{Introduction}
 
Geocoding of a location description is the process of transforming an address into a geographical coordinate. This process has been available in geographic information systems for quite some time \cite{Cooke1998} with applications for example for route planning,
validating customer addresses \cite{ARCGIS03}, or surveillance and management of disease outbreaks like the yearly wave of influenza \cite{Krieger:2002:Am-J-Epidemiol:12196317}.
However, with its ubiquitous use in modern web services (e.g., \cite{GoogleMaps,BingMaps}), requirements have become more severe: Since most of these services are free, geocoding servers must handle huge streams of queries at
very low cost. At the same time, users expect instantaneous answers. Finally, inputs will frequently be fragmentary, contain misspelled names or specify combinations of town and street that are inconsistent with the database. A service is likely to be more popular and useful if it tolerates such imprecisions.

While companies offering such services have naturally worked on this problem intensively, we are not aware of academic work that offers the required combination of low latency, high throughput, and error-correction for large address files. Our original aim was to make reasonable methods available to a cooperating company and to the
academic community. However, to our own surprise it turned out that our approach achieves better solution quality than the market leaders at low costs so that it might also help the industry to improve their services.

In this paper we focus on the algorithmic aspects of the problem to map information about town and street to a database entry for the intended street. We do not consider the problem
of resolving house numbers or producing highly accurate output coordinates.
Although this has been a major focus of previous literature, e.g., \cite{goldberg07:urisa}, we view it as orthogonal to our problem
since once town and street have been correctly identified, we are dealing with
much more local data and hence much smaller data volumes.  For example, searching an odd house number not present in
the database can often be done by a binary search in 
a sequence of the known odd house numbers followed by an interpolation \cite{drummond95,goldberg07:urisa}. On todays servers with many gigabytes of RAM, it might even be possible to precompute all estimated positions of house numbers.

The paper is structured as follows. Section \ref{sec:index} outlines the basic index data structure. 
Section \ref{sec:query} explains the query algorithm, while \ref{sec:rating} develops on a rating function to rank the matches from the query.
A variant of the algorithm for the case that the address is typed into a single field is explained in Section~\ref{sec:single-search-field}.
Section \ref{sec:neighborhood-graph} gives a data structure that resolves ambiguous town names both as a downstream computation step and in an online setting as real-time suggestions.
Section \ref{sec:experiments} reports on an experimental evaluation. Last but not least Section \ref{sec:conclusions} gives conclusions.

\subsection*{More Related Work}

Geocoded data used to cost several dollars per $1\,000$ records in the mid-eighties \cite{1566949} and didn't nearly provide the spatial accuracy of todays free services.
At that time the use of geographic information systems was limited to professionals only that were aware of the difficulties and limitations of the geocoding process \cite{geocoding-eval}.
In \cite{SGW08}, eight geocoders are compared. However, the evaluated Californian addresses are all given with high precision including city name, ZIP code, state and street name.

Improving the quality of geocoding using approximate string matching
is considered desirable in \cite{Winkler2004} but then dismissed as too expensive.

Approximate string matching itself has been studied intensively, 
e.g. \cite{1142879, Navarro00indexingmethods} but most actually implemented methods match only a specific pair of strings which is infeasible for large databases. Although there is considerable theoretical work on text indices 
allowing approximate matching \cite{Navarro00indexingmethods, 1142879}, there are few implementations because these methods are complicated and require superlinear space. 
Moreover, most previous work is on finding all matches in a single large text.
We will only require matching against full entries of a dictionary which is
a slightly simpler problem and thus might also allow more efficient solutions.
We build on one of the few available implementations \cite{Karch2010}, but still have to overcome the problem of superlinear space consumption.



Sengar \etal \cite{1341044, 1376745} describe a system that is able to handle ill-formed queries to a certain extent.
Their system does not require any country-specific rule set, but exploits the underlying geometric map data to produce a language independent representation of the data.
This kind of abstraction is especially useful in areas of the world where formal address formats are non-existing, e.g., in India.
However, it remains unclear how such systems scale to large databases.

\section{Index Data Structure}\label{sec:index}
Our basic input data are a set $T$ of \emph{towns} and a set $S$ of \emph{streets} that are fully defined by a name and a reference to a town. In this paper we use the term ``town'' both for 
a \emph{district} of another place and for an independent place. To avoid confusion
we will therefore use the term \emph{city} for independent places even if they are small.
Streets belonging to multiple towns are cut into pieces
belonging to a single town.
Districts additionally contain a reference to the city they belong to.

\subsection{Inverted Indices and Town Lists}
We view place and street names as (very short) documents containing a sequence of 
\emph{tokens} separated by white space, commas or hyphens. Thus we can use methods known from full text search to support fast geocoding. In particular, 
we build  two \emph{inverted indices}, i.e. the town index maps tokens appearing in town names to the towns using that token in their name and the street index maps tokens appearing in street names to all streets containing this token.

In addition to the above inverted index, we precompute the set $\towns(s)$
of town IDs containing a street with name $s$ and also a set $\towns(t)$ of 
town IDs with name $t$. This translation to town IDs will enable us to 
quickly determine which combinations of town and street name correspond
to actual addresses.

\subsection{Ignoring Light Tokens}
While indexing by token makes the index more convenient to use, it introduces a serious problem. 
A street query of the form \myText{New Hollywood Street} will return \emph{every} street that matches \emph{any} of the tokens \myText{New}, \myText{Hollywood}, or \myText{Street}.
In fact, about 25\% of all street names match the token \myText{Street}, therefore we would get a really big candidate set.
To avoid this problem, we adapt a concept which is often used in information retrieval and text mining \cite{ModernIR,1361686,872796}:
The \emph{inverse document frequency} of a token $c$ with respect to a set $M$
of strings (town or street names in our case) is defined as
$$\idf(c)\Is \log_2
\frac{\sum_{x\in M}|\tokens(x)|}%
     {|\setGilt{x\in M}{c\in\tokens(x)}|}$$
where $\tokens(x)$ is defined as the set of tokens making up string $x$.
Tokens that occur very often in the document (such as \myText{Street}) receive a lower IDF weight than those that appear only infrequently. 
Tokens that receive a high weight are more helpful in identifying the correct string, because they match fewer strings in the index. 

We will use these observations to our advantage:
When a user enters an address that they want to have geographically referenced, they may leave out parts of the address that they deem irrelevant, but they will probably enter those parts of the query that will non-ambiguously define what they are looking for.
In our example, the user may leave out either \myText{New} or \myText{Street}, but they most definitely won't leave out the token \myText{Hollywood}, which is also the token with the highest IDF weight among those three. 
If we expect the user to enter the most important part of an address, it is not necessary to have said address be referenced also by the remaining, unimportant tokens, i.e. we don't want to find the street \myText{New Hollywood Street} by the token \myText{Street}, because we expect that the more descriptive token \myText{Hollywood} will be entered anyway.
Let 
$$w_t^s := \frac{\idf(t)}{\sum_{t' \in \tokens(s)} \idf(t')}$$
be the \emph{relative weight} of the token $t$ in the string $s$. 
In our example the relative weights might be 0.31 for \myText{New}, 0.6 for \myText{ Hollywood}, and 0.09 for \myText{Street}, respectively. 
If we decide that the query must contain tokens that make up a fraction $\mu$ of the total weight, then we can ignore the lightest $k$ tokens, if the sum of their weights is not greater than $\mu$.
E.g., for $\mu > 0.4$, we can ignore the tokens \myText{New} and \myText{ Street}.
Note that this local definition of importance is very different from the stop words that are completely ignored in some inverted index data structures to save space. For example, suppose our database contains streets named \myText{Rhododendron Alley} and \myText{Alley Street} then it is likely that the index for
\myText{Alley} will contain \myText{Alley Street} but not \myText{Rhododendron Alley}.
This both save space, query time, and useless candidates. 
\dk{I would like to add the following example to emphasize the meaning of \emph{relative} weight: ps: removed as duplicate????}


\subsection{Approximate Token Indices}\label{sec:approx-token}
We also build indices supporting approximate search on the sets of \emph{tokens}
appearing in town names and street names respectively. 
This design decision has two crucial advantages over the more obvious choice to have an index on the town and street names themselves.
First, since the dictionaries are much smaller than the full data base, we can afford superlinear space to some extent. 
For example, our German input set contains 1.35 million streets 
but only 219 thousand distinct tokens for street names.
Furthermore, token based indices can easily handle queries that drop part of the town or street name. For example, most users just type \myText{Frankfurt} when they are looking for \myText{Frankfurt am Main}.
The approximate index \cite{Karch2010} for token set $M$ with maximum error $d_i$ can be queried with a string $q$ and returns a set $M_q\subseteq M$ of tokens that have edit distance (Levenshtein distance) at most $d_i$.

\section{Multi-Field Search}\label{sec:query}

We first concentrate on the case where a query consists of two strings typed into separate fields for town name and street name. Note that in this   case it is easy and largely orthogonal to allow additional fields for house   number or ZIP code.
\frage{dropped ablaufdiagramm for now. Reinclude if we have room and time to better adapt it to the paper.}
After normalization and tokenization (Section~\ref{sec:init}),
we try three increasingly sophisticated ways to obtain sets $\CT$ and $\CS$
of town and street \emph{candidates} respectively that allow an increasing number of errors (Section~\ref{sec:exact}--\ref{sec:approximate}).
After each of these attempts, we combine these candidate sets into
consistent candidate addresses from $\CT\times\CS$.
We stop as soon as we have found satisfying solutions.
When at the end no reasonable solutions have been found, an empty result is returned.

\subsection{Initialization Phase}\label{sec:init}

The input strings are first scanned and transformed into a set $Q$ of tokens and normalized to lower-case. This is also the place where some culture-specific preprocessing can be done. In our German implementation, there is only one such specialty: The German words for \myText{Street}, \myText{Lane}, \ldots are
sometimes used as a separate word and sometimes as a suffix and nobody really knows which version is correct in every case. Hence, compounds with these
suffixes are broken into a normal form with separate tokens.
This even works when the suffixes are misspelled or abbreviated.

\subsection{Partially Exact Town Match}\label{sec:exact}

Following the successful principle of ``make the common case fast'',
we use a simplified special treatment for the case of a \emph{partially exact town match} where at least one sufficiently
rare token of a town candidate is exactly matched. 
If this already yields a plausible result, we stop.
For example, in the query \myText{Franfrt am Main, R\"omerberg}, \myText{Frankfurt}
was misspelled. But the token \myText{Main} is an exact match and therefore we consider the set of towns that contain this token somewhere in their name before we look at all possible approximate matches.
If we find a street similar to \myText{R\"omerberg} in one of the candidates, we can stop.

\subsection{Periphery Search}\label{sec:periphery}

If we successfully identified partially exact town matches during the first phase, but could not match a street in these towns with a sufficient rating,
we extend the scope of exact search to the \emph{periphery}:
If the input specifies a city, we try all its districts, if it specifies a district, we try the city it belongs to and all its districts.
 \begin{figure}[h]
\begin{center}
   \includegraphics[width=0.5\columnwidth]{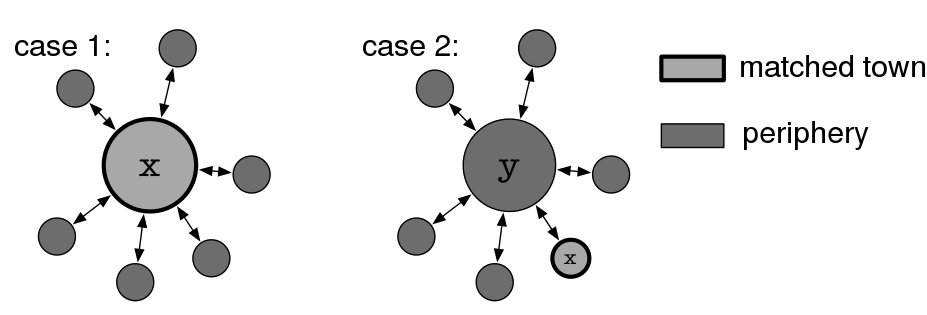}
\end{center}
 \caption{Two Cases of Periphery Search}
 \label{fig:periphery}
 \end{figure}

If the town provided by the user is matched against a candidate $x$ which is subsequently corrected to a town $y$ in the periphery of $x$, we still calculate the rating for $x$, because the name of $y$ generally does not match anything in the query string and would lead to a low rating.

\subsection{Approximate Search}\label{sec:approximate}

When there are no or no good partially exact matches or when even periphery search does not find a good candidate, additional
candidates are computed using the approximate indices for towns and streets. If a town candidate found specifies a district $x$ of a city $y$, we also add $y$ to the candidates.
However, we do not do a full scale approximate periphery search because this could
easily yield results that are hard to understand.

\subsection{Finding Compatible Candidates}
\label{sec:dropping-incompatible}


After partially exact matching, periphery search, or approximate search,
that all treat towns and streets separately, 
we generate address candidates where town and street are compatible with each other.
A pair $(t,s)\in \CT\times \CS$ is compatible
if a street with name $s$ is present in some town with name $t$, i.e.,
we have to compute the set 
$$\CST\Is\setGilt{(t,s)\in \CT\times \CS}{\towns(t)\cap\towns(s)\neq \emptyset}\punkt$$
There are various ways to do this more efficiently than the naive way of
computing $|\CT|\times|\CS|$ set intersections. Appendix~\ref{app:compatible} gives
one particularly efficient implementation.




\section{Rating Candidates}\label{sec:rating}


After we have dismissed most of the search space, we are left with a hopefully small set of compatible address candidates $(t,s) \in \mathcal{T}\times \mathcal{S}$. \frage{paragraph reformulated}These are then \emph{rated}.
The result is interpreted using two threshold values $\rlower$ and $\rupper$.
Ratings below $\rlower$ are considered unsatisfactory. If all results are unsatisfactory, more extensive search is done (after partially exact matching or 
periphery search) or, when everything failed, an empty result is returned.
In contrast, if a candidate with rating $\geq\rupper$ is found, the search
returns successfully without further attempts at refined searching.
Depending on the application we can then return the top ranked candidate or a list of good candidates.

To develop a rating heuristic, let us recall the different kinds of errors that we want to compensate for:
\begin{itemize}
\item Typing errors
\item Missing or redundant tokens
\item Inconsistent pairing of a street and a town.
\end{itemize}
Since periphery search and candidate filtering have already dealt with inconsistent candidates, we are left with the first two issues.

The first step on the way to a robust rating heuristic is to align the query to a candidate, i.e. find a good mapping from the tokens in the query to the tokens in the candidate (see Section~\ref{sec:matching-the-query-to-a-candidate}). Based on this mapping, we then compute the actual rating.

The rating is computed separately for town and street by the same method and combined by the arithmetic mean afterwards.
Hence, the following explanation details the town rating only.
\frage{new:}There is one small asymmetry however that we call \emph{filter by edit distance}: Since there are usually less candidate towns than candidate streets, we first filter out candidates that are already unsatisfactory because they do not sufficiently well fit the town description in the query.  

\subsection{Matching the Query to a Candidate}
\label{sec:matching-the-query-to-a-candidate}

\begin{figure}
  \centering
  \includegraphics[width=0.5\columnwidth]{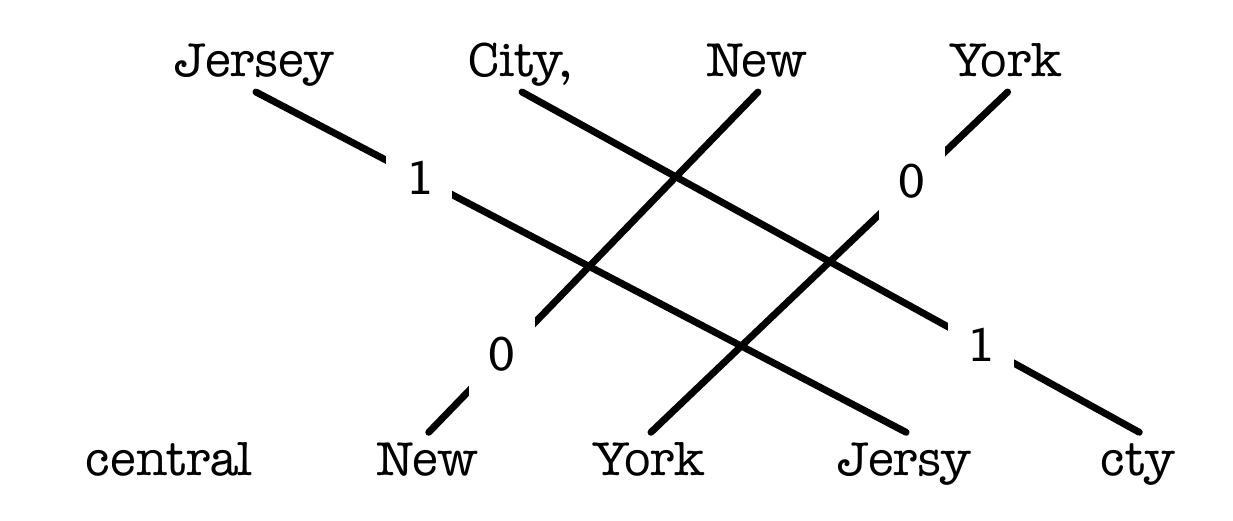}
  \caption{Candidate (top) is matched against query (bottom). Edge labels symbolize edit distances.}
  \label{fig:matching}
\end{figure}

To match the town tokens $q\in Q$ given by the users to the tokens of a candidate town name $c\in C$, 
we solve a \emph{minimum weight perfect matching problem} on a bipartite graph.
If $|Q|\leq |C|$ we add $|C|-|Q|$ \emph{dummy} nodes to $Q$ and obtain the matching graph $G=(Q\cup C,Q\times C)$ where
the weight of edge $(q,c)$ is the edit distance between $q$ and $c$
if $c$ is not a dummy node and $0$ if $c$ is a dummy node. 
Edit distances take misspellings into account and dummy query nodes go a long way to model missing tokens in the query.
Similarly, but perhaps less importantly,
if $|Q|> |C|$ we add $|Q|-|C|$ \emph{dummy} nodes to $C$. This matching problem can be solved in polynomial time (e.g. \cite{AhuMagOrl93,hungarian}). 
Moreover, the considered graphs are very small so that solutions can be computed quickly.
See Figure \ref{fig:matching} for an illustration.


\subsection{The Rating Function}
\label{sec:the-rating-function}
Based on the matching, we calculate a rating for each candidate. 
The rating should take into account the following considerations:
\begin{itemize}
\item 
  Each token that matches with at most $d$ errors should be awarded some points.
  It makes sense to choose $d$ larger than the error bound
  $d_i$ for the approximate index since space or index access time is
  no issue for the pairwise distance computations used for the rating function.
\item 
  Tokens that could not be matched with at most $d$ errors should not be awarded points and may even be punished.
\item 
  The user is more likely to omit information (either because they forget it or because they deem it unnecessary) than to over-specify the query.
  Therefore, tokens in the query that don't match anything in the candidate should be punished higher than candidate tokens that don't match anything in the query.
\item 
  The rating should be a real number in the interval $[0,1]$, 
  with one denoting a perfect match.
\item 
  The heuristic should be able to distinguish between tokens that are \emph{important} and tokens that do not provide much information.
\end{itemize}

Rather than directly using the edit distance,
we also want to take into account the lengths of compared words, since the rate of error that can be introduced into a word with a constant number of changes depends on its length.
We therefore define \emph{token similarity} between two tokens $q$ and $c$ as
$$  \mathrm{sim}(q,c) := 
  \begin{cases} 
    1-\frac{\ed(q,c)}{|c|}, & \mbox{if } \ed(q,c) \leq $d$\\ 
    0, & \mbox{else}
  \end{cases}
\punkt$$
We normalize the error rate by the length of $c$ since candidates are entries that are actually present in our database.

In order to take the \emph{importance} of a candidate token
into account, we use its inverse document frequency (see Section~\ref{sec:index}).

Let $M$ denote the set of edges $(d,c)$ between query tokens and candidate tokens that were matched with edit distance $\leq d$.
Let $U$ denote the set of unmatched query tokens, i.e., those
tokens that could not be matched to any candidate token with at
most $d$ errors.
We can now define our rating function
\begin{align*}
  \rating(Q,C)&\Is \gamma\rating^Q(Q,C)+(1-\gamma)\rating^C(Q,C)\\\text{ where}\\
  \rating^Q(Q,C)&\Is
\frac{
  \displaystyle\sum_{(q,c)\in M} (\simil(q,c))^{\alpha} \idf(c)
  }{
  \displaystyle\sum_{(q,c)\in M} \idf(c) + |U| \idf_{avg}
  }\text{ and}\\
  \rating^C(Q,C)&\Is
  \sum_{(q,c)\in M} \idf(c) / \sum_{c\in C} \idf(c)
\end{align*}
Where $\idf_{avg}$ is the average over the IDF-values of all town tokens.
The term $|M_c| \idf_{avg}$ expresses that the unmatched queries should have matched somewhere but we have no idea where -- so we use an average IDF-value.
The parameter $\alpha$ is used to adjust how important it is to have similar matches. 
Notice that $\rating^Q$ is not influenced by the number of unmatched \emph{candidate} tokens. This is why we
compute a convex combination of $\rating^Q$ with $\rating^C$ which penalizes unmatched candidate tokens.
The parameter $\gamma\in[0,1]$ specifies the relative weight.
Usually we want to give more weight to the matched parts of a query, therefore we choose $\gamma > 1/2$.

\section{Single-Field Search}
\label{sec:single-search-field}
In the previous sections we focused on separate fields for town and street 
because this interface is more important for our cooperation partner, because multi-field search is easier to program, and one should expect that it reduces errors.
From the users points of view, however, it is more convenient to enter a query into a single text field, with street and town in arbitrary order.
Online services like Google Maps or Bing Maps have therefore adopted single-field search.

However, in order to compare multi-field search and single-field search and
in order to compare our approach with internet services, we have also
implemented a simple version of single-field search with an emphasis on quality.

Our solution is based on the plausible hypothesis that the token sequence resulting from a single-field query has the format $\Id{streetToken}^*\Id{townToken}^+$ or $\Id{townToken}^+\Id{streetToken}^*$, i.e., street and town tokens are contiguous and there is at least one token designating a town.
We exhaustively try all $2m-1$ possible ways to split a token sequence of length $m$ and call a multi-field search for each of them. Refer to Section~\ref{sec:conclusions} for a discussion of possible optimizations.

For example, the query \myText{Oxford Street, London} can be split as
\begin{center}
\begin{tabular}{l|l}
Town   & Street        \\\hline
\tt Oxford & \tt Street London \\
\tt Oxford Street & \tt London \\
\tt Oxford Street London &\\\hline
\tt Street London & \tt Oxford\\
{\bf\tt\bf London} &{\bf\tt\bf Oxford Street} 
\end{tabular}
\end{center}
Only the last line would return a perfect rating as we might have hoped.
This query is a bit lucky though since there is no \myText{London Street} in \myText{Oxford} which, if it existed, would also receive a perfect rating.

\section{Neighborhood Graph}
\label{sec:neighborhood-graph}
In most countries, city names are not unique. 
For example, Wikipedia knows 41 Springfields, 5 of them in Wisconsin.
We present an efficient data structure here that 
can be used to find plausible nearby cities for disambiguation.
Applications could either use the data structure to propose 
disambiguating places or they could match them to inputs of the user
such as \myText{Springfield near Kansas City}. 

What makes a city plausible? It should be large, and it should be close to the city it is supposed to disambiguate. Hence, we are facing a bicriteria optimization problem. A safe way to handle such a situation is to consider all cities that are not dominated by any other city.  ($x$ dominates $y$ wrt city $t$ if it is both closer to $t$ than $y$ and larger than $y$.) This problem is known under several names: finding \emph{Pareto optima}, \emph{vector maxima} \cite{KIP75}, or a \emph{skyline} \cite{ICDE01*421}.  If the cities are sorted by size, it is easy to solve the problem with a full scan of all cities -- outputting a city if it is closer than all previously inspected ones. However, looking at all cities may still be too slow.

We will encode the required information into a graph.  Assume the cities are numbered 1 to $n$ by decreasing size.  
There is no need to store all towns, only those cities big enough 
to serve as a reference place.
A convenient way to define the size of a city here is to just count the number of streets. Let $D_i\Is(\set{1,\ldots,i}, E_i)$ denote the Delaunay triangulation\footnote{Note that for convenience we work with Euclidean geometry   here.  It is an interesting problem to consider what happens when we do   geometry on the surface of a sphere.} \cite{BKOS97} of the $i$ largest cities.

\sloppypar Then we will consider the \emph{neighborhood graph} $\NeG\Is(\set{1,\ldots,i},\cup_{i=1}^nE_i)$. Depending on the application, we may interpret $\NeG$ as a directed acylic graph where all edges go from smaller to larger indices (downward) or vice versa (upward). The intuition here is that the Delaunay triangulation encodes a natural concept of proximity. Directing the edges
\emph{upward}, i.e., towards larger cities allows us to find such cities.
Obviously, it is not enough to just consider the Delaunay triangulation $D_n$ of all $n$ cities since we would usually end up in a dead end of a medium sized cities whose neighbors are all smaller.
The following theorem states that the union $\NeG$ of $n$ Delaunay triangulation solves this problem very effectively:
\begin{figure}[b]
\begin{code}
\Procedure PareteOptima$(q)$\+\\
$u\Is 1$\RRem{start at top of hierarchy.}\\
\Repeat\+\\
  output $u$\\
  \Foreach neighbor $v$ of $u$ in $\NeG$\+\+\\ with $u<v$ in increasing order \Do\-\\
    \If $||q-\Id{position}(v)||_2 < ||q-\Id{position}(u)||_2$ \Then\+\\
      $u\Is v$\\
      break for loop\-\-\-\\
\Until no closer node found
\end{code}
\caption{Finding Pareto optimal cities in the neighborhood graph $\NeG$.\label{fig:pareto}}
\end{figure}

\begin{theorem}\label{thm:neighbor}
Consider any map position $(x,y)$. 
The Pareto optima with respect to city size and closeness to position $(x,y)$ forms a downward path from node $1$ to the city closest to $(x,y)$.
Moreover, this path can be found with the simple greedy algorithm depicted in Figure~\ref{fig:pareto}.
\end{theorem}
\begin{proof}
By induction from $1$ to $n$, analogous to the work of Birn \etal \cite{Birn2010}.
\end{proof}

Of course it is important how long it takes to construct $\NeG$ and how much space it takes. If the size of a city is assumed to be independent on its position, the problem is the same as in \emph{randomized incremental construction} of a Delaunay triangulation \cite{GKS92} -- we get a linear number of edges in time $\Oh{n\log n}$.

\section{Experimental Evaluation}\label{sec:experiments}

\paragraph*{Implementation}
We have implemented the system described above in C++ making extensive use of the STL library. This probably leaves open significant further tuning opportunities. For example, we use a simple merging based STL algorithm for computing set intersections rather than a tuned code as used in full text indices. The tuning parameters have been chosen intuitively without an attempt at finding optimal values: 
We ignore light tokens comprising up to 40 \% of the cumulative IDF of 
a name.
The correction limit of the approximate dictionaries and pairwise edit distance computations is limited to $d=d_i=2$ in order to keep space consumption low.  
Candidate matching use the Hungarian method \cite{AhuMagOrl93} using the implementation by \cite{Kuhn-MunkresAssignmentAlgorithm}.
In the rating function, similarities between matched words are taken to the power $\alpha=2$ and in the convex combination $\rating^Q(Q,C)=\gamma\rating^Q(Q,C)+(1-\gamma)\rating^C(Q,C)$ we choose $\gamma=3/4$.
The threshold for a satisfactory rating is $\rlower=1/2$ and a good rating
starts at $\rupper=4/5$.
Experiments were performed on an Intel Core i7 920@2.67GHz with 12GB RAM on Linux 2.6.27 using a single core.
The programs were compiled with GCC 4.3.2.

\paragraph*{Map Data}
We use commercial data (from 2009) comprising all German street and town names.
There are about 12\,000 cities, 
108\,000 towns, 80\,000 town names, 76\,000 town name tokens,
1\,350\,000 streets, 560\,000 of which contain the token \myText{Strasse},
444\,000 street names, and 269\,000 street name tokens.
A street name consists of 2.5 tokens on average, while town names consist of 1.1 tokens on average.
The input data takes about 30~MB space while our index data structures take
about 327 MB. A more frugal assignment of space to various hash tables
could reduce that to a bit more than 200 MB but we think that 327 MB is almost negligible for a server setting -- even very energy-efficient 32-bit servers based on Atom, or ARM could be used.

\subsection{Pseudorealistic Random Queries}
\label{sec:exp-random-queries}
For our experiments, we use a set of existing, \emph{relevant} addresses $R$, and a set of non-existing, \emph{irrelevant} addresses $I$.
A relevant address is sampled by first choosing a random street name $s$
and then picking a random town from $\towns(s)$.
An irrelevant is composed of randomly chosen town and street names
such that combination which accidentally occur in the database are rejected.
Ideally, we would like to return correct results for relevant address queries and no result for irrelevant address queries.
To generate a simple random query, it would be easiest to just insert, delete or substitute random characters in an existing address. 
The errors that are introduced this way, however, are unlikely to resemble the errors that a human would make while entering a query through a keyboard.
We identify several sources of errors to generate input sets with more realistic errors.

Typing errors are very common and we distinguish 
\begin{itemize}
\item swapped characters
 
  \emph{Example:} \myText{Frankfurt} $\rightarrow$ \myText{Frankfrut}

\item missing characters

  \emph{Example:} \myText{Frankfurt} $\rightarrow$ \myText{Franfurt}

\item superfluous or wrong characters, mostly closely located to the correct character on the keyboard (here in terms of the German \textsc{QWERTZ}-layout).

  \emph{Example:} \myText{Frankfurt} $\rightarrow$ \myText{Frankdfurt} or \myText{Frankdurt}
\end{itemize}

Depending on the respective language there are several sources of error that are phonetic.
\begin{itemize}
\item doubled characters where there should be a single character

\emph{Example:} \myText{Dublin} $\rightarrow$ \myText{Dubblin}

\item single character where there should be two of the same

\emph{Example:} \myText{Cardiff} $\rightarrow$ \myText{Cardif}

\item The \textsc{Soundex} algorithm identifies classes of characters such that different characters from the same class differ only slightly in their pronunciation.

\emph{Example:} \texttt{z} $\equiv$ \texttt{s}, \myText{Zaragoza} $\rightarrow$ \myText{Saragosa}

\item Two consecutive vowels that occur in the same syllable are called a \emph{diphthong}. 
In German, for example, several different diphthongs sound the same or similar:
\begin{itemize}
\item $\mathtt{ei} \equiv \mathtt{ey} \equiv \mathtt{ay} \equiv \mathtt{ai}$
\item $\mathtt{eu} \equiv{} $\texttt{\"{a}u} $\equiv \mathtt{oy} \equiv \mathtt{oi}$
\item \dots
\end{itemize}

\emph{Example:} \texttt{Hoyerswerda} $\rightarrow$ \texttt{Heuerswerda}
\end{itemize}

In order to introduce $k$ errors into an address, we introduce $\lceil k/2 \rceil$ errors into the street string and $\lfloor k/2 \rfloor$ errors into the town string. To introduce an error, we first pick a random error class,
then a random token, and then a random position. All distributions are uniform.

Here we report experiments on 1\,000 relevant and 100 irrelevant addresses.
We classify the results returned by the index as follows:
\begin{itemize}
\item \emph{True Positive (TP):} A relevant address that is correctly identified.
\item \emph{True Negative (TN):} An irrelevant address that does not return a result or a correct partial result (i.e. the correct town).
\item \emph{False Positive (FP):} An irrelevant address where the index does return a result.
\item \emph{False Negative (FN):} A relevant address where we don't find a result.
\item \emph{Incorrectly Identified (II):} A relevant address that returns an incorrect result, i.e. another relevant address.
\end{itemize}
The match rates, along with the query times, are shown in Table~\ref{tab:match-rates}.

Multi-field search works extremely well, both with respect to result quality and query time. Only at five errors, when three errors are introduced into the street name, we see a sharp increase of false negative results. At this point, the approximate street index will often fail to find the right result because its error limit is set to $d_i=2$.
Interestingly, query times \emph{decrease} with the number of errors.
The reason is that we have to check a smaller number of candidates both in
the approximate index and when rating candidates.

Single-field search for relevant addresses works almost as well as multi-field search. The only noticable difference is that a small fraction of the 
false negative results mutates into incorrectly identified results. For irrelevant addresses, our current implementation seems to be too aggressive though because it returns a significant number of false positives. 
On the first glance it looks paradoxical that we get a larger number of output errors when there are no spelling errors in the input. But the reason is simple: without spelling errors we obtain higher ratings for the generated candidates and thus it becomes more likely that the result is accepted. This indicates that
the result quality could be improved by increasing the threshold for accepting a result. Single-field query times are an order of magnitude larger than for multi-field search. This is not surprising, since our current implementation
naively factors a single-field search into several multi-field searches.

\begin{table}
  \centering
  \begin{tabular}{rr|r|r|r|r|r}
      \multicolumn{7}{c}{{\bf Multi-field search}}\\
      \cline{2-6}
    & \multicolumn{3}{|c|}{relevant} & \multicolumn{2}{|c|}{irrelevant} & \\
      \hline
    \multicolumn{1}{|r|}{\textbf{Errors}} & \textbf{TP} & \textbf{FN} & \textbf{II} & \textbf{TN} & \textbf{FP} & \multicolumn{1}{|r|}{\textbf{Time [ms]}} \\
    \hline
     \multicolumn{1}{|r|}{0}& 1\,000 & 0 & 0  & 93 & 7& \multicolumn{1}{|r|}{3.02}  \\
     \hline                   
     \multicolumn{1}{|r|}{1}& 989  & 10 & 1 & 95 & 5& \multicolumn{1}{|r|}{2.75} \\
     \hline                   
     \multicolumn{1}{|r|}{2}& 988  & 11 & 1 & 94 & 6& \multicolumn{1}{|r|}{2.44}  \\
     \hline                   
     \multicolumn{1}{|r|}{3}& 928  & 66 & 6 & 94 & 6& \multicolumn{1}{|r|}{2.40}  \\
     \hline                   
     \multicolumn{1}{|r|}{4}& 854 & 140 & 6 & 99 & 1& \multicolumn{1}{|r|}{1.79} \\
     \hline                   
     \multicolumn{1}{|r|}{5}& 557& 431 & 12 & 97 & 3& \multicolumn{1}{|r|}{1.59} \\[3mm]
    \hline
      \multicolumn{7}{c}{{\bf Single-field search}}\\
        \cline{2-6}
    & \multicolumn{3}{|c|}{relevant} & \multicolumn{2}{|c|}{irrelevant} & \\
      \hline
    \multicolumn{1}{|r|}{\textbf{Errors}} & \textbf{TP} & \textbf{FN} & \textbf{II} & \textbf{TN} & \textbf{FP} & \multicolumn{1}{|r|}{\textbf{Time [ms]}} \\
    \hline
    \multicolumn{1}{|r|}{0} & 1\,000 & 0 & 0   & 52 & 48 & \multicolumn{1}{|r|}{26.07}\\
    \hline                                         
    \multicolumn{1}{|r|}{1} & 989 & 10 & 1   &63 & 37 &  \multicolumn{1}{|r|}{23.33}\\
    \hline                 
    \multicolumn{1}{|r|}{2} & 986 & 13 & 1   &74 & 26 &  \multicolumn{1}{|r|}{19.72}\\
    \hline                 
    \multicolumn{1}{|r|}{3} & 927 & 66 & 7   &75 & 25 &  \multicolumn{1}{|r|}{18.44}\\
    \hline                 
    \multicolumn{1}{|r|}{4} & 856 & 125 & 19 &80 & 20 &  \multicolumn{1}{|r|}{16.69}\\
    \hline                 
    \multicolumn{1}{|r|}{5} & 560 & 414 & 26 &86 & 14 &  \multicolumn{1}{|r|}{14.31}\\
    \hline
  \end{tabular}
  \caption{
    Matching rates and query times for random addresses -- 
    1\,000 relevant ones and 100 irrelevant ones.
  }
  \label{tab:match-rates}
\end{table}

\begin{figure*}
  \centering
  \includegraphics[width=0.49\textwidth]{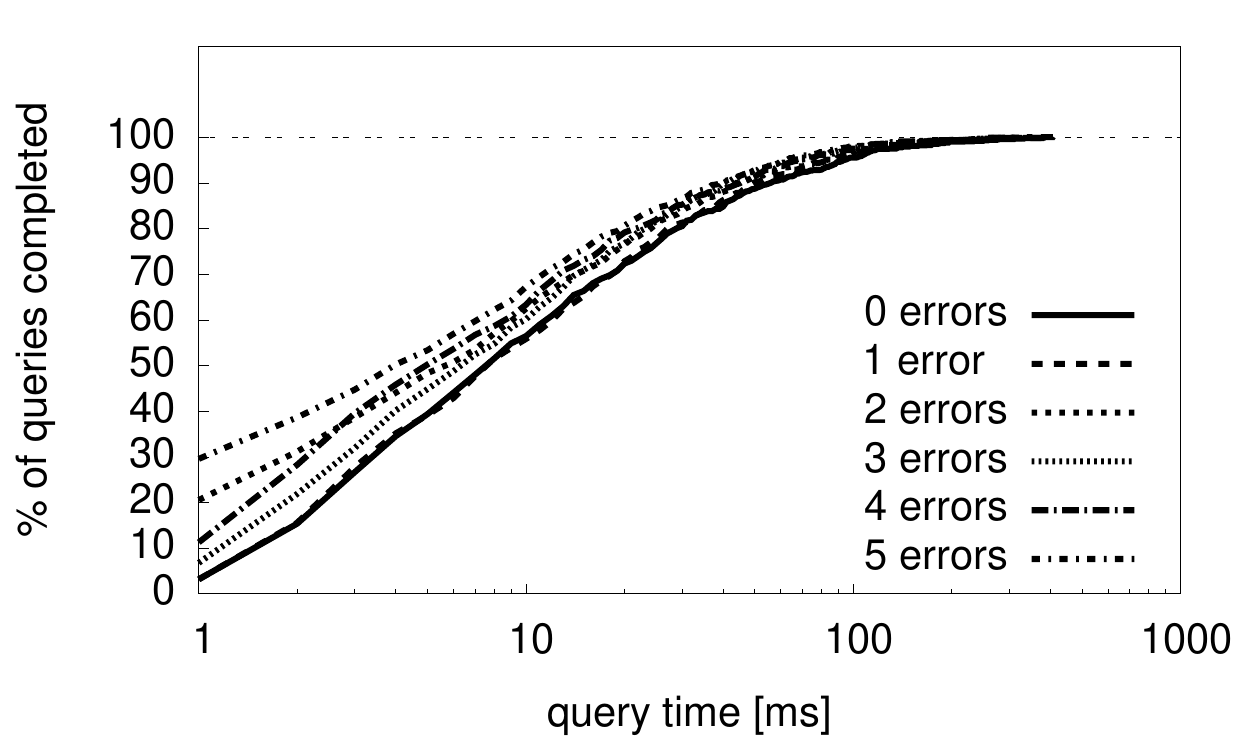}
  \includegraphics[width=0.49\textwidth]{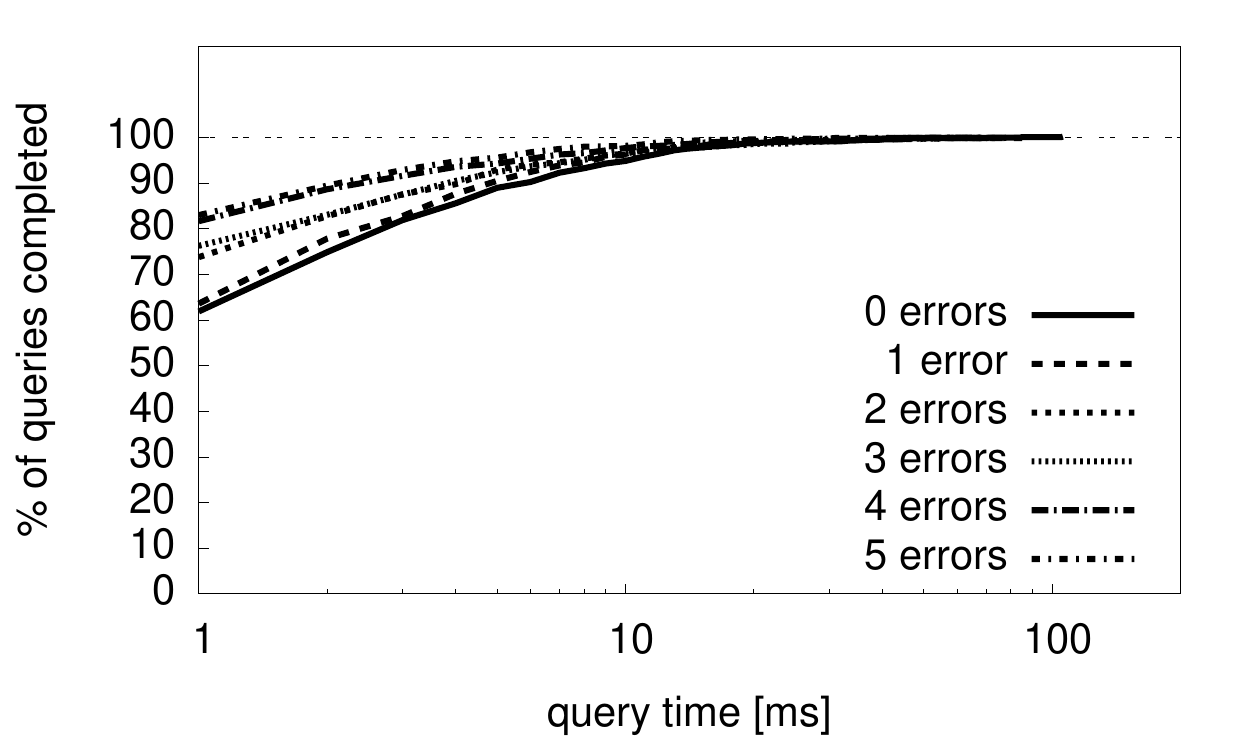}
  \caption{
   Query times of single-field (left) and multi-field (right) search for several numbers of distortions.
  }
  \label{fig:runtime-plot}
\end{figure*}

Figure \ref{fig:runtime-plot} shows the distribution of query times for the same set of $5\times 1\,100$ queries. 
$90\%$ of all multi-field queries finish in less than $5$ ms. The maximum
query time observed is 106 ms. Note that for the server scenario we are considering, we want very low \emph{average} query time to achieve high throughput
and low cost. Occasional slower queries are no problem as long as they 
do not lead to noticable delays for the user. 100 ms is well below the delays
users are accustomed to experience due to network latencies anyway.
Although single-field search is an order of magnitude slower on the average, this slow-down does not translate into a proportional increase of the slowest query times -- we still remain below 406 ms.
This indicates that the query times of the generated multi-field subqueries are
not strongly correlated.

\begin{table}
  \centering\begin{tabular}{|l||r|r|r|}
    \hline
    $d$&\textbf{Bing}&\textbf{Google}&\textbf{Ours} \\
    \hline\hline
      0 & 87 & 97 & 100\\
\hline      1 & 78 & 54 &  98\\
\hline      2 & 71 &  3 &  98\\
\hline      3 & 59 &  3 &  94\\
\hline      4 & 55 &  2 &  86\\
    \hline
  \end{tabular}
\caption{number of resolved queries}
\label{tab:api-compare}
\end{table}

We also compared against existing geocoders.
To do so, we took the first 100 relevant queries from the above query set and ran them against the publicly available APIs%
\footnote{It should be noted that the subjective performance of Google with interactive use on Google-Maps is much better than over the API. In particular, the auto-completion mechanism works very well but is obviously not applicable to the API.} of Bing and Google Maps.
Table \ref{tab:api-compare} reports on the number of true positive results.
Google works very well for undistorted inputs -- the three remaining errors could perhaps be attributed to differences in the input data. But already for a single error, the recognition rate drops to 54 \% and completely collapses for $d\geq 2$. 
Bing already has significant deficits at $d=2$ but
fails more gracefully for distorted inputs. Still, the number of failed answers is an order of magnitude larger than for our system.

\subsection{Real-world Queries}
\label{sec:exp-real-queries}

To get an impression of the quality of the results, we also did experiments with real-world input, i.e. actual queries that have been provided by users of an existing geocoder\footnote{Complete references will be given in the final version of paper.}.
The test data consists of multi-field 1\,383 queries that were logged
from the the users of a cooperating company.
The results were pre-classified by the company into five categories.
We briefly describe the categories.

\textbf{Exact: }
Queries where each token can be matched without errors to a token in the result.
The differences allowed between query and result are those that are handled by a normalization phase.
\\[.1cm]
\emph{Example}: \myText{london tally road} $\rightarrow$ \myText{London, Tally Road}

\textbf{Partially Exact: }
Queries where each token occurs in the result, but not each token of the result string occurs in the query.
\\[.1cm]
\emph{Example}: \myText{london tally} $\rightarrow$ \myText{London, Tally Road}

\textbf{High/Medium/Low: }
Queries that contain errors. 
The labels \Id{High}, \Id{Medium}, and \Id{Low} were assigned depending on the ``confidence'' of the old geocoder to have a correct interpretation of the input.
\\[.1cm]
\emph{Example}: \myText{Lodon; Tall Rd} $\rightarrow$ \myText{London, Tally Road} (e.g. classified as \Id{Medium} )

We tested our address search with these queries checking 
correctness of our result manually.
We classify the results into three categories:

\textbf{Strong Match: }
A result that is unquestionably correct. 
In many cases the query was non-ambiguous and easy to verify.

\textbf{Weak Match: } 
A result that is not correct, but has successfully identified parts of the query, e.g. the town.

\textbf{No Match: }
Either a result that is definitely incorrect, or no result at all.

In total, the test data consists of 1\,383 queries, classified into 844 Exact, 357 Partially Exact, 125 High, 41 Medium, and 16 Low queries.
Since they had to be verified by hand, we picked random samples of at most 100 queries per class.
Also, we removed those that were not resolvable (e.g. for shopping centers, water parks etc. -- our index does not contain points of interest).
There remained 100 exact and 99 partially exact queries, as well as 81, 34 and 15 queries classified as high, medium and low.
The results are shown in Table \ref{tab:real-queries}.

\begin{table}
  \centering
\scalebox{0.91}{
  \begin{tabular}{lr||r|r|r||r|r|r||r|r|r|}
    \cline{3-11}
    & & \multicolumn{3}{|c|}{\textsc{\textbf{Bing}}} & \multicolumn{3}{|c|}{\textsc{\textbf{Google}}} & \multicolumn{3}{|c|}{\textsc{\textbf{Ours}}} \\
    \hline
    \multicolumn{1}{|l|}{\textbf{Class}}   & \textbf{\#} & \textbf{s} & \textbf{w} & \textbf{n} & \textbf{s} & \textbf{w} & \textbf{n} & \textbf{s} & \textbf{w} & \textbf{n} \\
    \hline
    \multicolumn{1}{|l|}{{Exact}}   & 100 & 81 &  5 & 24 & 100 &  0 & 0 & 100 & 0 & 0\\
    \hline
    \multicolumn{1}{|l|}{{Partial}} &  99 & 77 & 20 &  2 &  96 &  2 & 1 &  99 & 0 & 0\\
    \hline
    \multicolumn{1}{|l|}{{High}}    &  81 & 60 & 16 &  5 &  65 & 10 & 6 &  77 & 2 & 2\\
    \hline
    \multicolumn{1}{|l|}{{Medium}}  &  34 &  7 &  1 & 26 &  16 & 11 & 7 &  27 & 6 & 1\\
    \hline
    \multicolumn{1}{|l|}{{Low}}     &  15 &  1 &  1 & 12 &   9 &  4 & 2 &  13 & 1 & 1\\
    \hline
  \end{tabular}
}
  \caption{The match rates of our geocoder on real-world queries.}
  \label{tab:real-queries}
\end{table}

Our system outperforms the Google and Bing API in all categories. As with the real world inputs, Google is similarly good for (partially) exact queries but looses ground for errorneous inputs. For example, for category \emph{High} our system returns correct results for all but 4 of the inputs whereas Google fails for 16 inputs -- a factor four in the failure rate.
An interesting difference to the random inputs is that now Google consistently outperforms Bing -- also for the inputs with errors.

\subsection{Parameters that affect the Query Time}
\label{sec:query-time-address-index}

We use several techniques to make sure that the number of candidates stays small and that we don't have to perform too many edit distance computations.
To see if these techniques are necessary and how each of them affects the query time, we have performed a number of of experiments.
The techniques are:

\textbf{Filter Incompatible Candidates (FIC): }
As described in Section \ref{sec:dropping-incompatible}, we keep only those town and street candidates that are geographically compatible.

\textbf{Filter by Edit Distance (FED): }
As described in Section \ref{sec:rating} we have an additional filtering stage
before full rating evaluation the drops candidates that can already eliminated
because the town names are an unsatisfactory match.

\textbf{Ignore Light Tokens (ILT): }
As described in Section~\ref{sec:index}, we can ignore some tokens during the construction of the index due to their weight in comparison to the other tokens.
E.g. the candidate \myText{New Hollywood Street} will be represented only by \myText{Hollywood}, because the other two tokens occur so frequently in the dictionary that they would not be of much help to distinguish this candidate from others.

\begin{table}
  \centering
  \begin{tabular}{|c|c|c|r|}
    \hline
     \textbf{ILT}   &\textbf{FIC}   & \textbf{FED}   &\textbf{Query Time [ms]} \\
            \hline\hline
     \texttimes &\texttimes & \texttimes &  570.00 \\
            \hline
     \texttimes &\texttimes & \checkmark  &  566.00 \\
            \hline    
     \texttimes &\checkmark  & \texttimes &  199.00 \\
            \hline
     \texttimes &\checkmark  & \checkmark  & 126.00 \\
            \hline
     \checkmark  &\texttimes & \texttimes &  10.68 \\
            \hline
     \checkmark  &\texttimes & \checkmark  & 10.58 \\
            \hline
     \checkmark  &\checkmark  & \texttimes & 3.45 \\
            \hline
     \checkmark  &\checkmark  & \checkmark  &  2.09\\
    \hline
  \end{tabular}
  \caption{The effect of the features ILT, FIC and FED on the lookup time.}
  \label{tab:features}
\end{table}

As we can see in Table \ref{tab:features}, disabling ILT absolutely destroys the performance.
To see why this is so, consider the most frequent token in the street dictionary, \myText{street}. 
If we randomly choose any street, the probability that it contains the token \myText{street} is about $1/3$.
Without ILT, \emph{any} query that contains the token \myText{street} will return \emph{all} candidates that contain this token.
Hence, if we randomly choose a candidate and query the index with this candidate, we can expect a candidate set that contains at least $1/9$ of all streets, which is almost 50\,000 candidates for our data.
In our experiments the actual number of candidates was even bigger, 67\,000 candidates on average.
The query time drops by a factor of almost 100 when we enable ILT.
FIC gives us another boost of factor 3 and FED makes a difference only when used in conjunction with FIC.
None of these features has a noteworthy effect on the memory requirements of the index, therefore all of them should be enabled by default.

\subsection{Neighborhood Graph}
\label{sec:experiments-neighborhood-graph}

We have build a neighborhood graph for all 12\,379 \emph{cities} in our input, 
resulting in 71\,896 edges, i.e., there are less than 6 edges per node which is 
similar to what we would expect for random city sizes. About 290 nodes are reachable from a city on the average, saving a factor $>40$ compared to a full
scan of the city table, even if we do not fully use Theorem~\ref{thm:neighbor}.

We also performed a small user study to test our model.  We randomly picked 48 towns from a map of Germany and asked a non-expert in the field of geography or computer science to attribute a larger reference town to the one we picked. In 46 of these cases, a Pareto-optimal reference town was chosen.

\section{Conclusions and Future Work}\label{sec:conclusions}
We presented algorithms and data structures for error-correcting geocoding 
that yield instantaneous answers at costs negligible compared to
the overheads for displaying maps answers, etc. over the internet.
Perhaps most surprising is that we still have a high match rate with as much
as four errors and this is much better than current web geocoders. 
Another pleasant surprise was that our combination of powerful data structures and general rating functions yields a considerably simpler solution than several rule based industrial solutions we have heard about.

Although we are already quite fast, we still have significant tuning
opportunities. In particular,
it will be relatively easy to further speed up single-field search.
For example, we currently do not even cache accesses to the approximate dictionary or results edit distance computations. More generally, faster algorithms for edit distance computation could be tried \cite{Hyyro2005}.

Although our experiments have so far focused on commercial German road data, we believe that they are easy to adapt to other Western industrial countries.
In particular, these countries have
similar address systems and language conventions. Points of interest (POI) like gas stations or restaurants can be incorporated by treating them like a street, town or house number depending on the context. Finding street intersections is relatively easy once we have geocoded the two intersecting streets.

The basic ingredients -- fast approximate dictionary search, token matching and scoring functions might also help in other settings like in countries with more complicated addresses or less structured reference data. However in that case we should expect more errors, longer query times, and the need for further heuristics.

Our concept of neighborhood graph is a promising approach to disambiguate
queries involving frequent town names. One interesting aspect is that
the set of nodes reachable through the neighborhood graph $\NeG$ may contain further ``interesting'' nodes. In particular, we may want to restrict the set of reported cities to those in the same ``region'' (e.g., county, state, or nation)
as the query town. Since Delaunay triangulations encode neighbors ``in all directions'', we suspect that with appropriately defined ``locally well behaved region shapes'' the desired output still consists of  nodes reachable in
$\NeG$.
 
\bibliographystyle{abbrv}
\bibliography{references,geocoder,bibliography}
\begin{appendix}
\section{Fast Computation of \texorpdfstring{$\CST$}.}\label{app:compatible}

We describe this algorithm in an appendix since it is an application of folklore tricks in maintaining sets of small integers and thus not a really new result.
On the other hand, the applications of these tricks may not be obvious in this case so that a description is sensible.
We first compute the union $\towns(\CS)$ of all sets $\towns(s)$ for $s\in \CS$.
$\towns(\CS)$ is represented as an array $A$ with one entry for each town ID.%
\footnote{A is initialized only once during program startup. Afterwards it suffices to clean entries that have actually been used. To facilitate this, the
used town IDs are stored in a stack.}
While doing this, we annotate each entry of $\towns(\CS)$ with a list $\ell$ of pointers to the street names that enter this town into the union.
Then, for each $t\in\CT$ and each $x$ in $\towns(t)$ we remember $\setGilt{(t,s)}{s\in A[x].\ell}$ as compatible candidates. 
The algorithm is linear in the size of
the inspected lists $\towns(t)$, $\towns(s)$, and the output size.

\end{appendix}
\end{document}